\documentclass[preprint2]{emulateapj}

\slugcomment{Accepted to ApJL on April, 6th 2012}
\usepackage{natbib,hyperref}
\begin{document}

\title{Detection of Thermal Emission from a Super-Earth}

\author{Brice-Olivier Demory\altaffilmark{1}, Micha{\"e}l Gillon\altaffilmark{2}, 
Sara Seager\altaffilmark{1,3}, Bjoern Benneke\altaffilmark{1}, Drake Deming\altaffilmark{4} and Brian Jackson\altaffilmark{5}}

\altaffiltext{1}{Department of Earth, Atmospheric and Planetary Sciences, Massachusetts Institute of Technology, 77 Massachusetts Ave., Cambridge, MA 02139, USA. demory@mit.edu}
\altaffiltext{2}{Institut d'Astrophysique et de G\'eophysique, Universit\'e de Li\`ege, All\'ee du 6 Ao\^ut, 17, Bat. B5C, Li\`ege 1, Belgium.}
\altaffiltext{3}{Department of Physics and Kavli Institute for Astrophysics and Space Research, MIT, 77 Massachusetts Avenue, Cambridge, MA 02138, USA.}
\altaffiltext{4}{Department of Astronomy, University of Maryland, College Park, MD 20742-2421, USA}
\altaffiltext{5}{Carnegie Institution of Washington, Department of Terrestrial Magnetism, 5241 Broad Branch Road NW, Washington, DC, 20015, USA.}

\begin{abstract}
We report on the detection of infrared light from the super-Earth 55\,Cnc\,e, based on four occultations obtained with \textit{Warm Spitzer} at 4.5\,$\mu$m. Our data analysis consists of a two-part process. In a first step, we perform individual analyses of each dataset and compare several baseline models to optimally account for the systematics affecting each lightcurve. We apply independent photometric correction techniques, including polynomial detrending and pixel-mapping, that yield consistent results at the 1-$\sigma$ level. In a second step, we perform a global MCMC analysis including all four datasets, that yields an occultation depth of $131\pm28$\,ppm, translating to a brightness temperature of $2360\pm300\,K$ in the IRAC-4.5\,$\mu$m channel. This occultation depth suggests a low Bond albedo coupled to an inefficient heat transport from the planetary dayside to the nightside, or else possibly that the 4.5\,$\mu$m observations probe atmospheric layers that are hotter than the maximum equilibrium temperature (i.e., a thermal inversion layer or a deep hot layer). The measured occultation phase and duration are consistent with a circular orbit and improves the 3-$\sigma$ upper limit on 55\,Cnc\,e's orbital eccentricity from 0.25 to 0.06.
\end{abstract}


\section{Introduction}

The nearby sixth magnitude naked-eye star 55\,Cnc is among the richest exoplanet systems 
known so far, with five planetary companions detected since 1996 \citep{Butler:1997, Marcy:2002, McArthur:2004, Wisdom:2005, Fischer:2008, Dawson:2010}. 
The recent discovery of the transiting nature of 55\,Cnc\,e, made independently in the visible 
with the \textit{MOST} satellite \citep{Winn:2011a} and in the infrared with the \textit{Spitzer 
Space Telescope} \citep[][hereafter D11]{Demory:2011}, set this super-Earth among the most 
promising low-mass planets for follow-up characterization.

A recent data reanalysis combining fifteen days of \textit{MOST} monitoring and two \textit
{Spitzer} transit observations, allowed us to refine 55\,Cnc\,e's properties \citep[][hereafter G12]
{Gillon:2012} that result in a planetary mass of $M_p = 7.81 \pm 0.56\: M_\oplus$  and a 
planetary radius of $R_p= 2.17 \pm 0.10\: R_\oplus$. Although they do not uniquely constrain 
the planetary composition, the mass and radius (and hence density) are consistent with a solid 
planet without a large envelope of hydrogen or hydrogen and helium. The planet could be a 
rocky core with a thin envelope of light gases. 
Another possible mass/radius interpretation could be an envelope of supercritical water 
above a rocky nucleus, where the exact amount of volatiles would depend on the composition 
of the nucleus (G12). In this scenario, 55\,Cnc\,e would be a water world similar to the core of Uranus and Neptune.  

The super-Earth size of 55\,Cnc\,e together with an extremely high equilibrium temperature ranging between 1940 and 2480\,K, motivated us to apply for \textit{Spitzer} Director's Discretionary Time to search for the occultation of 55\,Cnc\,e at 4.5\,$\mu$m. The goal was two-fold. First, the occultation depth provides an estimate of the brightness temperature, constraining both the Bond albedo and the heat transport efficiency between the planetary day and night-side \citep[e.g.,][]{Cowan:2011b}. Second, a measurement of the occultation phase and duration provides a constraint on a potential non-zero orbital eccentricity \citep{Charbonneau:2003} that could be maintained by the interactions with the four other planets of the system. 

We present in this Letter the first detection of light from a super-Earth. The new \textit{Warm Spitzer} 55\,Cnc observations and corresponding data reduction are presented in Section~2, 
while the photometric time-series analysis is detailed in Section~3. We discuss in Section~4 
the implications for our understanding of this planet, especially regarding the 
constraints on its atmospheric properties and orbital evolution.

\section{Observations}

Four occultation windows were monitored by {\it Spitzer} in the 4.5\,$\mu$m channel of its IRAC 
camera \citep{Fazio:2004a} in January 2012. Table~1 presents the description of each 
Astronomical Observation Request (AOR). For each occultation, 6230 sets of 64 subarray 
images were acquired with an individual exposure time of 0.01s. All the data were calibrated 
by the {\it Spitzer} pipeline version S19.1.0 which produced the basic calibrated data (BCD) 
necessary to our reduction. For all runs except the first one, the new Pointing Calibration and 
Reference Sensor (PCRS) peak-up mode\footnote{\url{http://irsa.ipac.caltech.edu/data/SPITZER/docs/irac/pcrs_obs.shtml}} was enabled. Because of the intrapixel sensitivity
variability of the 
IRAC InSb detectors coupled to the point response function\footnote{We followed here the IRAC instrument handbook terminology: \url{http://irsa.ipac.caltech.edu/data/SPITZER/docs/irac}} (PRF) undersampling, the measured 
flux of a point source shows a strong correlation with the intrapixel position of the star's center. 
This well documented ``pixel-phase effect'' creates a correlated noise that is the main limitation 
on the photometric precision of \textit{Warm Spitzer} \citep{Ballard:2010b}. The new PCRS 
peak-up mode aims at mitigating this correlated noise by improving the telescope pointing's 
accuracy.

We apply the same reduction procedure for all AORs. We first convert fluxes from the 
{\it Spitzer} units of specific intensity (MJy/sr) to photon counts. We then perform aperture 
photometry on each subarray image using the \textsc{APER} routine from the IDL Astronomy 
User's Library\footnote{\url{http://idlastro.gsfc.nasa.gov/contents.html}}. We compute the stellar 
fluxes in aperture radii ranging between 2.0 and 4.0 pixels, the best results being 
obtained with an aperture radius of 3 pixels for the first and fourth AOR, 2.8 pixels for the 
second AOR and 2.9 pixels for the third AOR. We use background annuli extending from 11 
to 15.5 pixels from the PRF center. The center and full width at half maximum (FWHM, along $x$ and $y$ axes) of the PRF is measured by fitting a Gaussian profile on each image using the \textsc{MPCURVEFIT} 
procedure \citep{Markwardt:2009}.

Examination of the background time-series reveals an ``explosion'' of the flux in the third and fourth AORs, with a similar shape to the one observed in D11. These two background ``explosions'' seem to be correlated with an increase of the measured FWHM along the $x$-axis. Because of the ``pixel-phase'' effect, changes in the PRF's shape create an additional correlated noise contribution (Gillon et al., in prep.).

For each block of 64 subarray images, we discard the discrepant values for the 
measurements of flux, background, $x$-$y$ positions and FWHM using a 10-$\sigma$ median clipping for the six parameters. We then average the resulting values, the photometric errors being taken as the uncertainties on the average flux measurements. At this stage, a 50-$\sigma$ clipping moving average is used on the resulting light curve to discard totally discrepant subarray-averaged fluxes. The number of frames kept for each AOR are shown in Table~1.

Figure~1 shows the raw lightcurve, the background flux, the $x$-$y$ centroid positions and FWHM for each AOR. The improved stability of the telescope pointing achieved by the new PCRS peak-up mode can be easily noticed for the $x$ time-series but seems less sharp along the pixel's $y$ axis.

\begin{figure*}[h!]
\begin{center}
\plotone{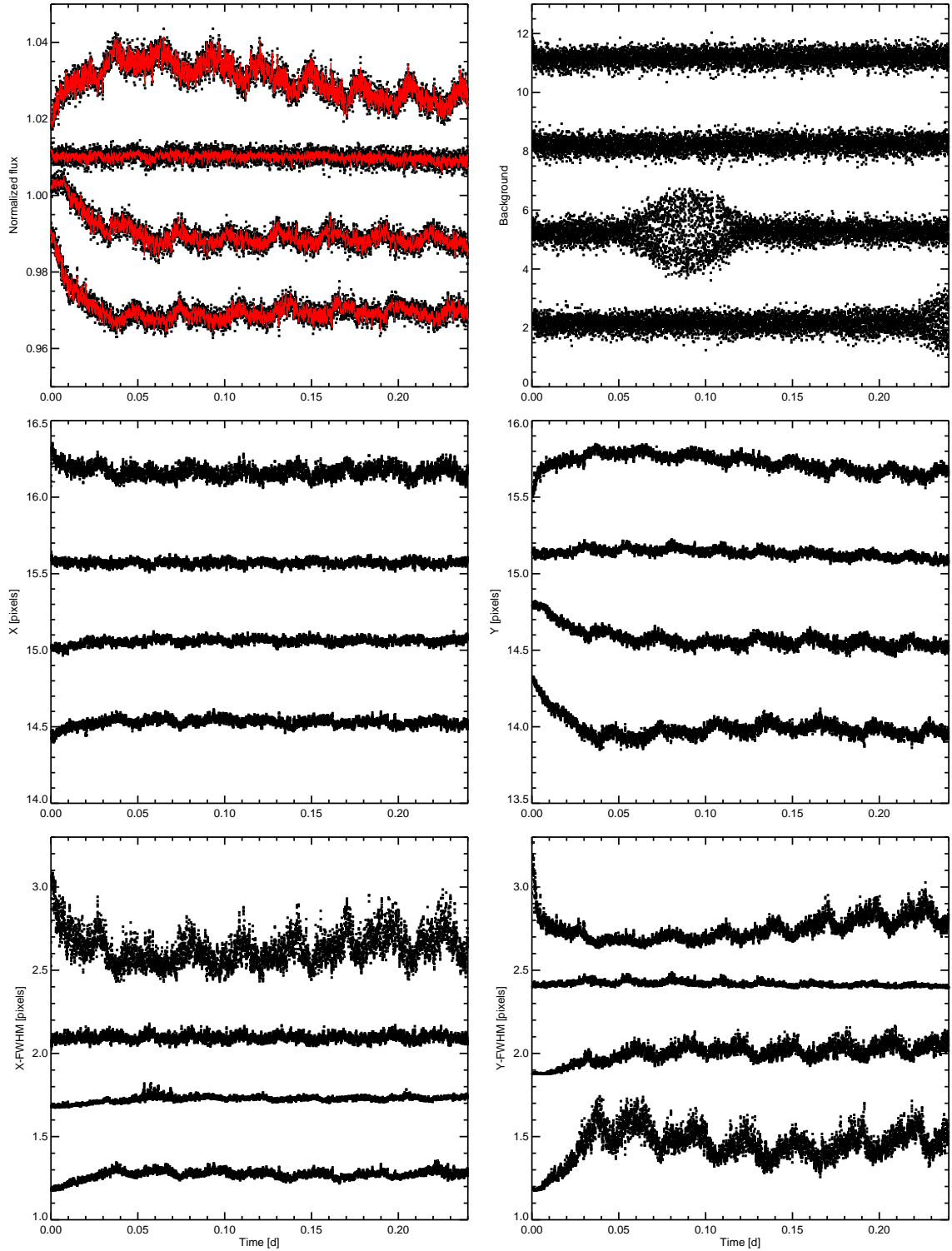}
\caption{Flux with best-fit baseline models superimposed (top left), background (top right), centroid position on the x-axis (middle left) and y-axis (middle right), PRF FWHM on the x-axis (bottom left) and y-axis (bottom right), from first (top) to fourth AOR(bottom). The time-series are arbitrarily shifted vertically for clarity.
\label{fig1}}
\end{center}
\end{figure*}

\section{Data Analysis}

\subsection{Independent analysis of each AOR}

The aim of this step is to perform an exhaustive model comparison to determine the optimal 
baseline model for each AOR. The baseline model accounts for the time- and position-
dependent systematic effects relevant to the IRAC-4.5\,$\mu$m observations (see D11, G12 
and references therein). For this purpose, we individually analyze each AOR by employing 
our adaptative Markov-Chain Monte Carlo (MCMC) implementation described in \citet{Gillon:2010a}. We set the occultation depth as a jump parameter and fix the orbital period $P$, 
transit duration $W$, time of minimum light $T_0$ and impact parameter $b=a \cos i/R_{\star}$ 
to the values obtained from the global analysis of the system (G12). We further impose a 
Gaussian prior on the orbital eccentricity ($e=0.06\pm0.05$, D11), allowing the eclipse phase 
and duration to float in the MCMC. For each model, we run two chains of $10^4$ steps each. 
Throughout this work, we assess the convergence and good mixing of the Markov chains using the statistical test from \citet{Gelman:1992}.

We first assume a baseline model based on a classical second order $x$-$y$ position 
polynomial (D11, eq.~1) to correct the ``pixel-phase'' effect, added to a time-dependent linear 
trend. We then increase this basic baseline complexity by trying combinations of up to third- 
and fourth-order $x$-$y$ position polynomials, second-order logarithmic ramp models and 
second-order time-dependent polynomials. The baseline models tested are therefore 
characterized by 7 to 20 free parameters, well constrained by the $\sim$6200 measurements 
of each AOR. We finally compute the Bayesian information criterion \citep[BIC, e.g.,][]{Gelman:2003} for all combinations and choose, from the MCMC output, the baseline model 
that yields the highest marginal likelihood. We show the resulting individual lightcurves on 
Fig.~2 and the selected model along with the occultation depth for each AOR in Table~1.

The correlated noise affecting each lightcurve is taken into account following \citet{Gillon:2010a}. A scaling factor $\beta$ from the comparison of the standard deviation of binned and 
unbinned flux residuals is determined during a preliminary MCMC run. This factor is then 
applied to the individual uncertainties of the flux time-series. The $\beta$ values for each AOR 
are shown in Table~1. To obtain an additional estimation of the residual correlated noise, we conduct a residual-permutation bootstrap analysis on each lightcurve corrected from the baseline model, similar to D11. This part of the analysis yields parameters in good agreement with the results from our MCMC analyses, albeit with significantly smaller error bars, suggesting that the error budget is dominated by the uncertainties on the coefficients of the complex baseline model and not by the residual correlated noise contribution.

The examination of the individual lightcurves obtained with more complex models lead to 
similar results to the lightcurves shown on Fig.~2, that are obtained with the baseline models selected in the previous step (Table~1). The resulting occultation depths are 
compatible within 1$\sigma$, securing both our detection and the insensitivity of the occultation 
signal to our adopted method for systematics correction.

As in D11 and G12, we perform a Lomb-Scargle periodogram analysis \citep{Scargle:1982} 
on the residuals of our selected model for each AOR. Peaks with marginal significance are 
found at 69\,min in the second AOR and at 51\,min in the third AOR, while neither of the first nor 
fourth AOR's residuals reveal a periodic signal. We include these sinusoidal modulations 
in the baseline models of our second and third AOR and perform a new MCMC that yields 
a higher BIC value than our model selected during the previous step of the analysis. We 
therefore neglect the sinusoidal term in both AORs. No hint of a $\sim50$\,min and
$\sim100$\,ppm amplitude modulation similar to the one reported in D11 and G12 is found in the four datasets.

We further build a ``pixel map'' to characterize the intrapixel variability on a fine grid. This 
approach has been already demonstrated by \citet{Ballard:2010b} and \citet{Stevenson:2011a} as an efficient method to remove the flux modulation due to the ``pixel-phase'' effect. The improved tracking accuracy brought by the new PCRS peak-up mode is indeed expected to increase the resolution of the pixel map, hence motivating this step of the analysis for our observations. More specifically, our implementation divides the area covered by the PRF in a grid made of 30$\times$30 boxes and counts the number of out-of-eclipse datapoints that fall in a each box. If a given box has at least four datapoints spanning at least 50\% of the AOR duration, the individual fluxes are divided by their mean, otherwise the corresponding measurements are rejected. As this method could average out an eclipse signal located in the designated out-of-eclipse parts of the lightcurve, we repeat this procedure by gradually shifting the out-of-eclipse datapoints in time. On average, as compared to the first AOR (obtained without the PCRS peak-up mode), the measurements sample 24\% more time per box in the second AOR, 57\% in the third AOR and 48\% in the fourth AOR. The occultation depth and duration obtained using the pixel-map corrected lightcurves are in good agreement with the ones employing a polynomial baseline model.

The different analysis techniques applied above yield detrended time-series in which 
the occultation signal is visible by eye with consistent phase and duration in three out of the 
four AORs (Fig.~2). The second AOR is the only one in which the detection is 
marginal ($68\pm52$\,ppm). The different approaches used to correct the photometry from the 
``pixel-phase'' effect show that the intrapixel sensitivity does not explain this discrepancy. We 
notice that the $\beta$ factor for this second AOR is $\sim$50\% larger than 
for the other AORs (see Table 1), suggesting a larger amount of correlated noise of 
instrumental or astrophysical origin. Examination of the onboard temperature sensors 
readings and other external parameters in the FITS files do not reveal any unusual patterns. 
While an actual variability of the planet's emission cannot be ruled out, the marginal 
disagreement (1.2$\sigma$) of the second occultation's amplitude relative to the three others 
is probably caused by this larger correlated noise.

\begin{deluxetable*}{lcccccc}
\tabletypesize{\scriptsize}
\tablecaption{Individual AOR and global fit properties}
\tablenum{1}
\tablehead{\colhead{} & \colhead{AOR\,1} & \colhead{AOR\,2} & \colhead{AOR\,3} & \colhead{AOR\,4} & \colhead{GLOBAL (prior on $e$)} & \colhead{GLOBAL ($e$ fixed)}}
\startdata
\tableline
& \\
Observation date [UT] &  2012-01-18  & 2012-01-21 & 2012-01-23 & 2012-01-31 & --- & --- \\
Observation window [UT] & 02:45 - 08:38 & 01:16 - 07:09 & 06:24 - 12:17 & 08:56 - 14:49 & --- & --- \\
Number of measurements & 6187 & 6190 & 6133 & 6164 & --- & --- \\
Baseline model\tablenotemark{a}  & $xy^4 + t^2 + r^2$ & $xy^2 + t^2$ & $xy^4 + t^2 + r^2$ & $xy^2 + t^2 + r^2$ & --- & --- \\
$\beta$ factor\tablenotemark{b}  & 1.25 & 1.89 & 1.20 & 1.27 & --- & --- \\
Occultation depth [ppm] & $202\pm54$ & $68\pm52$ & $107\pm53$ & $187\pm56$ & $131\pm28$ & $122\pm21$ \\
$\sqrt{e}\cos \omega$ & $0.037^{+0.054}_{-0.042}$ & $ -0.072^{+0.162}_{-0.108}$ &  $0.036^{+0.086}_{-0.077}$ & $0.001^{+0.061}_{-0.044}$ &  $0.094^{+0.025}_{-0.019}$ & ---\\
$\sqrt{e}\sin \omega$ & $0.003^{+0.079}_{-0.081}$  & $0.008^{+0.151}_{-0.165}$ &  $-0.001^{+0.092}_{-0.100}$ &  $0.010^{+0.092}_{-0.088}$ &  $0.001^{+0.028}_{-0.030}$ & ---\\
\enddata
\tablenotetext{a}{Baseline models are described by position ($xy$), time ($t$) and ramp ($r$) terms, with the order indicated in superscript.}
\tablenotetext{b}{See Section~3 for details on the $\beta$ scaling factor.}
\end{deluxetable*}

\begin{figure*}
\begin{center}
\epsscale{0.9}
\plottwo{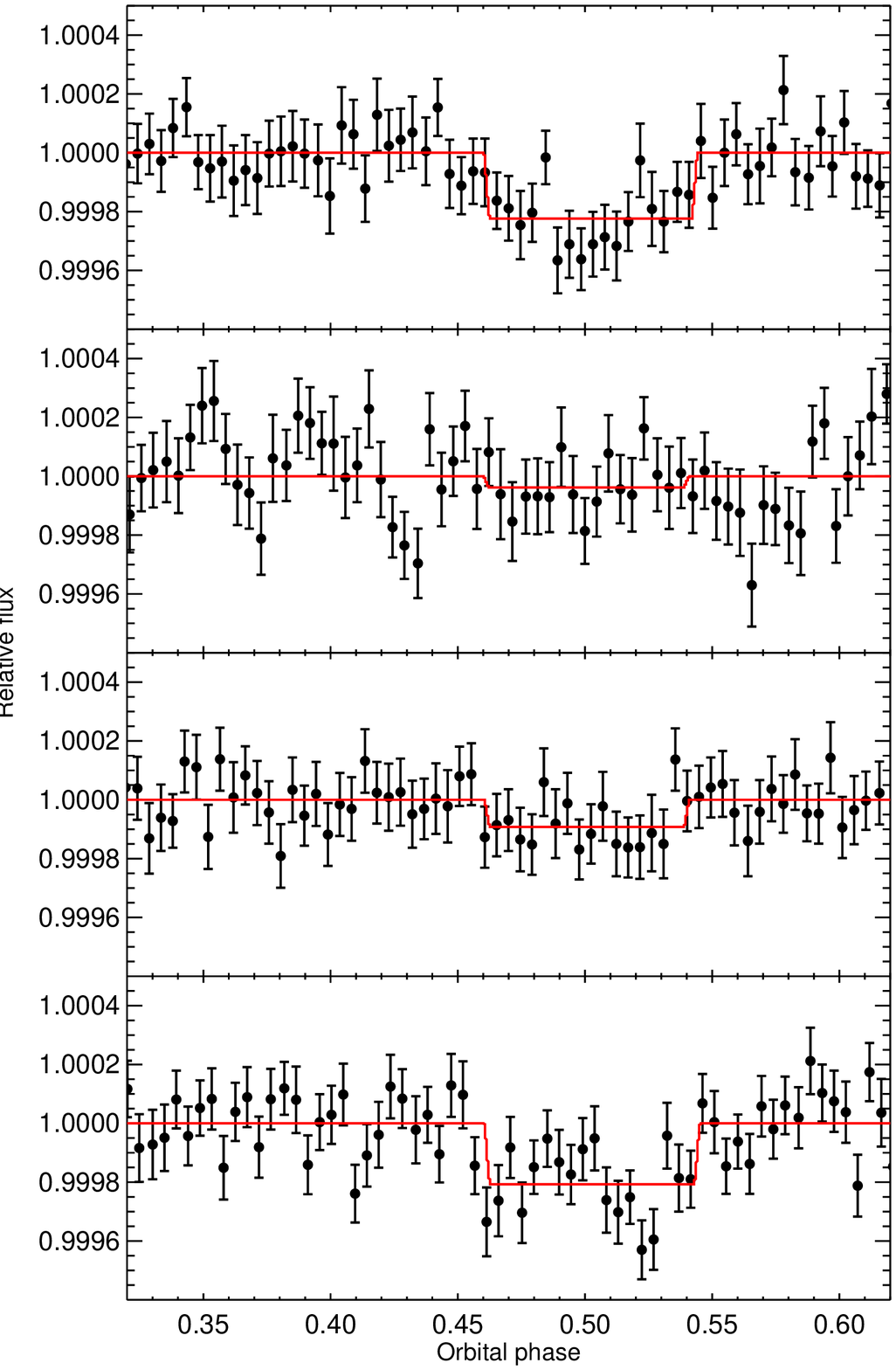}{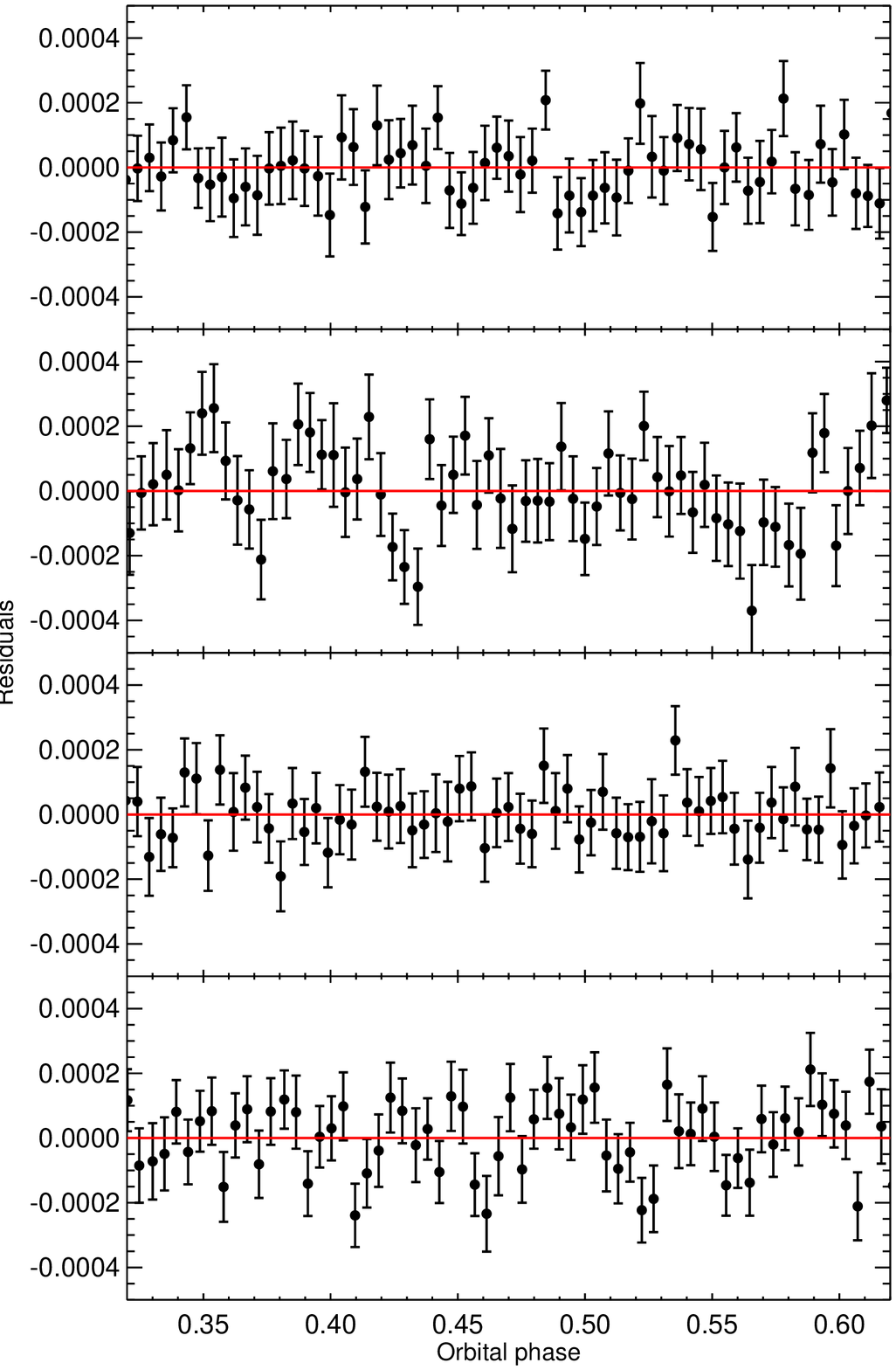}
\caption{Left: lightcurves obtained during the independent analysis of each AOR (see Sect.~3) 
divided by the best-fit baseline models. Right: residuals for each AOR. AORs are displayed from first (top) to fourth AOR (bottom). Lightcurves are binned per 5 min. Individual detections in each AOR are consistent both in phase and duration. 
\label{fig2}}
\end{center}
\end{figure*}

\subsection{Global analysis}

The final step of our analysis consists of performing a global analysis, including all four 
lightcurves in the same MCMC framework, to constrain both the occultation depth and orbital 
eccentricity. 
For this purpose, we assume Gaussian priors on $b$, $W$, $T_0$, $\sqrt{e} \cos \omega$, 
$\sqrt{e} \sin \omega$ and the stellar parameters based on the posterior distributions derived in D11, G12 and in \citet{von-Braun:2011a}. 
We first run two MCMC (one with the occultation model and one without) to assess the robustness of our detection, using as input the four raw lightcurves. We employ the baseline models selected in the previous step and shown in Table~1. These two runs are composed of three chains of $10^4$ steps each. 

We find an occultation depth of $131\pm28$\,ppm, and an eccentricity $e < 0.06$ 
(3-$\sigma$ upper limit). The odds ratio computed using the BIC between the two models (with 
vs. without occultation) is $\sim 10^4$ in favor of the occultation model. The thermal emission 
from 55\,Cnc\,e is therefore firmly detected.

To test the robustness of the eccentricity signal, we then perform an identical MCMC run but 
with the assumption of a circular orbit. The odds ratio between the circular and non-circular 
orbits is $\sim 10^3$ in favor of the circular case. The resulting occultation depth in this 
case is $122\pm21$\,ppm, in excellent agreement with the value obtained for the non-circular 
case. The phase-folded occultation lightcurve with the best-fit circular model is shown on Figure 3.

\begin{figure*}
\begin{center}
\epsscale{0.9}
\plotone{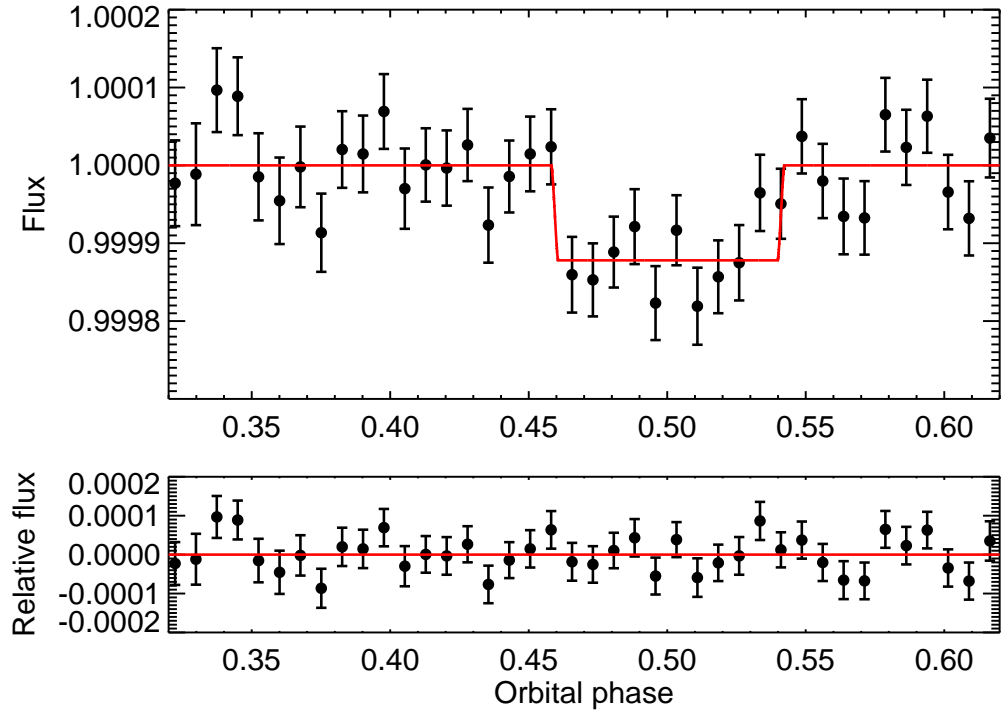}
\caption{Phase-folded occultation lightcurve resulting from our global analysis presented in 
Sect.~3. The best-fit circular model is superimposed in red. The lightcurve is binned per 8 min. 
\label{fig3}}
\end{center}
\end{figure*}

\section{Results and Discussion}

\subsection{Planetary Properties}

From the measured secondary eclipse depth we obtain a brightness temperature
estimate of $2360\pm300$\,K, using a stellar blackbody emission spectrum, with $T_{eff}=5196\pm24$\,K \citep{von-Braun:2011a}. At face value, the high brightness
temperature suggests that either 55\,Cnc\,e has both a low Bond albedo
and an inefficient heat transportation from the day side to the night
side, or the observations in the IRAC $4.5\,\mu m$ bandpass probe layers
in the atmosphere that are at temperatures higher than the equilibrium
temperature (Figure 4). Either scenario may explain why the brightness
temperature is observed to be higher than the zero Bond albedo
equilibrium temperature, $T_{eq}=1950$\,K, for a uniformly
re-radiating planet.

One interpretation could be that the planet is a rock with only a
minimal atmosphere established through vaporization \citep{Schaefer:2009,Leger:2009,Castan:2011}. Rocky objects
in the Solar System, e.g., Mercury and the Moon, have low Bond albedos
between 0.07 and 0.12. Lacking a thick atmosphere, a rocky super-Earth
also does not have any efficient mean of transporting heat from the
day-side to the night-side. We disfavor the bare rock scenario, however, 
because the 55\,Cnc\,e mass and radius measurements (D11, G12) 
exclude a rocky composition similar to Mercury
and Earth; to be a rocky planet with minimal atmosphere 55\,Cnc\,e would have to 
have the unlikely bulk composition of pure silicate.

Alternatively, if 55\,Cnc\,e has a substantial gas atmosphere or envelope,
the scenario of inefficient heat redistribution is still supported by the $T=2360$\,K brightness 
temperature, 
as long as the radiation at 4.5\,$\mu$m is not coming from a very hot layer, such as a deep 
layer or a
thermal inversion layer. The interpretation of 55\,Cnc\,e having a 
supercritical water envelope above a solid nucleus (D11) fits with the inferred high 
temperature;  
the absence of clouds at such a high temperature results in a water envelope with
a naturally low albedo. Although we cannot fully exclude the case of heat redistribution,
for the water planet or even a gas atmosphere over a rocky core, the probe to deep hot layers 
or the thermal inversion would have to be extreme. A probe of extremely deep, hot atmosphere
layers is unlikely because many of the gases likely to be
present in the atmosphere, in particular $\mbox{CO}_{2}$ and $\mbox{CO}$,
have high absorption cross sections in the spectral region of the
IRAC bandpass between 4 and 5\,$\mu$m. A thermal inversion, while present in Solar System 
planet atmospheres and 
suggested to be present in many similarly highly-irradiated hot Jupiters
\citep[e.g.,][]{Burrows:2007b,Knutson:2008}, is also an extreme explanation because the 
temperature would have to be
$\sim500$\,K above most of the rest of the atmosphere.  In general, while we know the total 
amount of energy re-radiated by the planet,
it is the interplay between absorption of stellar radiation at short
wavelength, the opacities at wavelengths at which the bulk radiation
is emitted ($\sim$1-5\,$\mu$m for a body at T$\sim2000$\,K), and
the opacity in our IRAC 4.5\,$\mu$m bandpass that sets the observed
brightness temperature. 

Considering the observational uncertainty, the observed brightness
temperature in the IRAC bandpass could be as low as $1830$\,K at the 2-$\sigma$
level. At this level of uncertainty, we can say  that
at least one or more of the following statements are true: 1) the
planet's Bond albedo is low; 2) the planet has an inefficient heat
transport from the day to night side; and/or 3) the IRAC 4.5\,$\mu$m
bandpass probes at atmospheric levels that are considerably hotter
than the day-side equilibrium temperature. 

\begin{figure*}
\begin{center}
\epsscale{0.8}
\plotone{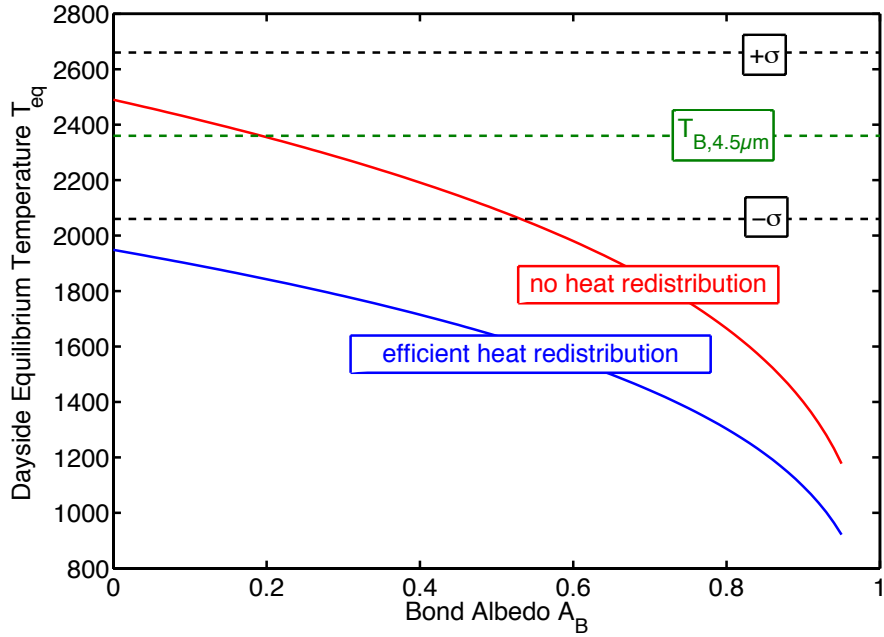}
\caption{Comparison of day-side equilibrium temperature with observed brightness
temperature in IRAC 4.5\,$\mu$m channel. Unless the IRAC 4.5\,$\mu$m
observations probe extremely hot layers in the atmosphere of 55 Cnc
e, the observed high brightness temperature at 4.5\,$\mu$m favors
weak or no heat redistribution and a Bond albedo $A_B < 0.5$. At 2-$\sigma$
uncertainty, both scenarios with either efficient heat redistribution
or no heat redistribution are in agreement with the observations. \label{fig4}}
\end{center}
\end{figure*}

\subsection{Orbital Eccentricity}

Our global MCMC analysis improves the 3-$\sigma$ upper limit on 55\,Cnc\,e's orbital eccentricity from 0.25 (D11) to 0.06.
However, even such a small eccentricity is unlikely because tidal interactions
between the planet and star would probably damp an initial eccentricity that large in a
few Myrs. Much larger initial eccentricities would be damped on similar
timescales. However, high order dynamical interactions among the
planets in the system may maintain an eccentricity for 55\,Cnc\,e of a
few parts per million, by comparison with theoretical studies of 
multi-planet systems having close-in
super-Earths \citep{Barnes:2010}. Strong observational
constraints on 55\,Cnc\,e's eccentricity may even provide information
about the planet's tidal response and internal structure
\citep{Batygin:2009a}. While a small eccentricity might have
little direct influence on observation, the concomitant tidal
dissipation within the planet can have dramatic geophysical
consequences, perhaps powering vigorous volcanism and resupplying the
planetÕs atmosphere \citep{Jackson:2008a}. The planet's
proximity to its host star suggests, in fact, it may be shedding its
atmosphere, and so active resupply may be necessary for long-term
atmospheric retention.

\subsection{Future prospects}

From the first, landmark detection of infrared light from a hot Jupiter \citep{Charbonneau:2005, Deming:2005} and further occultation observations of more than two-dozen Neptune- and Jupiter-sized planets \citep[see, e.g.,][]{Deming:2009b}, \textit{Spitzer} remains today the best instrument for exoplanet high-precision near-infrared photometry. At the dawn of the super-Earth sized planet discovery era \citep[see, e.g.,][]{Batalha:2012}, \textit{Spitzer's} continued legacy shows the need for keeping this observatory operational.  With the future launch of the \textit{James Webb Space Telescope}, the exoplanet community anticipates thermal emissions measurements for a number of different super Earths \citep[see, e.g.,][]{Deming:2009c}, hence improving our knowledge of this class of planets in a way similar to the achievements made by \textit{Spitzer} for hot Jupiters.

\acknowledgments
We thank Julien de Wit, Renyu Hu, Sean Carey and Nikole Lewis for insightful discussions and the anonymous referee for a report that improved the paper.
This work is based in part on observations made with the \textit{Spitzer Space Telescope}, 
which is operated by the Jet Propulsion Laboratory, California Institute of Technology under a 
contract with NASA. We thank the \textit{Spitzer} Science Center staff and especially Nancy 
Silbermann for the efficient scheduling of our observations. M. Gillon is FNRS Research 
Associate.

{\it Facilities:} \facility{Spitzer}


\end{document}